\documentclass[preprint,prb,aps,showpacs,footinbib,amsmath,amssymb,superscriptaddress]{revtex4-1}
\usepackage{graphicx}
\usepackage{dcolumn}
\usepackage{color}
\usepackage{bm}
\usepackage{natbib,hyperref}
\usepackage{amsmath,amssymb,amsfonts,bbold}
\usepackage[normalem]{ulem}

\newcommand{\beq}{\begin{equation}}
\newcommand{\eeq}{\end{equation}}
\newcommand{\beqa}{\begin{eqnarray}}
\newcommand{\eeqa}{\end{eqnarray}}

\begin{document}

\title{Scanning tunneling spectroscopy investigations \\ of superconducting--doped topological insulators: \\
		Experimental pitfalls and results}

\author{Stefan Wilfert}
\email[corresponding author:\ ]{swilfert@physik.uni-wuerzburg.de}
	\address{Physikalisches Institut, Experimentelle Physik II, 
		Universit\"{a}t W\"{u}rzburg, Am Hubland, 97074 W\"{u}rzburg, Germany}
\author{Paolo Sessi} 
	\address{Physikalisches Institut, Experimentelle Physik II, 
		Universit\"{a}t W\"{u}rzburg, Am Hubland, 97074 W\"{u}rzburg, Germany}
\author{Zhiwei Wang}
	\address{Physics Institute II, University of Cologne, 50937 Cologne, Germany}	
\author{Henrik Schmidt}
	\address{Physikalisches Institut, Experimentelle Physik II, 
		Universit\"{a}t W\"{u}rzburg, Am Hubland, 97074 W\"{u}rzburg, Germany}
\author{M.\ Carmen Mart{\'i}nez-Velarte}
	\address{Department of Quantum Nanoscience, Kavli Institute of Nanoscience, 
		Delft University of Technology, Lorentzweg 1, 2628 CJ Delft, The Netherlands}
\author{Seng Huat Lee} 
\altaffiliation{Current address: 2D Crystal Consortium, Materials Research Institute, 
		The Pennsylvania State University, University Park, Pennsylvania 16802, United States}
	\address{Department of Physics, Missouri University of Science and Technology, 
		Rolla, MO 65409, USA}
\author{Yew San Hor}
	\address{Department of Physics, Missouri University of Science and Technology, 
		Rolla, MO 65409, USA} 
\author{Alexander\,F.\,Otte}
	\address{Department of Quantum Nanoscience, Kavli Institute of Nanoscience, 
	Delft University of Technology, Lorentzweg 1, 2628 CJ Delft, The Netherlands}
\author{Yoichi Ando}
	\address{Physics Institute II, University of Cologne, 50937 Cologne, Germany}
\author{Weida Wu} 
	\address{Physikalisches Institut, Experimentelle Physik II, 
		Universit\"{a}t W\"{u}rzburg, Am Hubland, 97074 W\"{u}rzburg, Germany}	
	\address{Department of Physics and Astronomy, Rutgers University, 
		Piscataway, New Jersey, 08854, USA}
\author{Matthias Bode} 
	\address{Physikalisches Institut, Experimentelle Physik II, 
	Universit\"{a}t W\"{u}rzburg, Am Hubland, 97074 W\"{u}rzburg, Germany}	
	\address{Wilhelm Conrad R{\"o}ntgen-Center for Complex Material Systems (RCCM), 
	Universit\"{a}t W\"{u}rzburg, Am Hubland, 97074 W\"{u}rzburg, Germany}    
	   


\date{\today}

\begin{abstract}
Recently the doping of topological insulators has attracted significant interest 
as a potential route towards topological superconductivity. 
Because many experimental techniques lack sufficient surface sensitivity, however, 
a definite proof of the coexistence of topological surface states 
and surface superconductivity is still outstanding.   
Here we report on highly surface sensitive scanning tunneling microscopy (STM) 
and spectroscopy (STS) experiments performed on Tl-doped Bi$_2$Te$_3$, 
a three-dimensional topological insulator which becomes superconducting in the bulk at $T_{\rm C} = 2.3$\,K.  
Landau level spectroscopy as well as quasiparticle interference mapping clearly demonstrated 
the presence of a topological surface state with a Dirac point energy $E_{\textrm{D}} = -(118 \pm 1)$\,meV 
and a Dirac velocity $v_{\textrm{D}} = (4.7 \pm 0.1)\cdot 10^{5}$\,m/s. 
Tunneling spectra often show a superconducting gap, but temperature- and field-dependent measurements 
show that both $T_{\rm C}$ and $\mu_0 H_{\rm C}$ strongly deviate from the corresponding bulk values.  
Furthermore, in spite of a critical field value which clearly points 
to type-II superconductivity, no Abrikosov lattice could be observed.  
Experiments performed on normal-metallic Ag(111) prove 
that the gapped spectrum is only caused by superconducting tips, 
probably caused by a gentle crash with the sample surface during approach.  
Nearly identical results were found for the intrinsically n-type compound Nb-doped Bi$_2$Se$_3$.
Our results suggest that the superconductivity in superconducting-doped V-VI topological insulators 
does not extend to the surface where the topological surface state is located.  
\end{abstract}


\maketitle

\newpage

\section{Introduction}
  

It has quite early been theoretically recognized that the lack of spin- and spatial-rotation symmetries 
in $p$-wave superconductors leads to unconventional textures of the order parameter 
which may result in domain walls and quasiparticle excitations with vanishing excitation energies, 
so-called zero modes.\cite{Volovik1989,Volovik1992}
Whereas early theories were originally designed to describe superconductors 
in symmetries characteristic for the fractional quantum Hall effect (FQHE), 
the discovery of three-dimensional topological insulators (TIs),\cite{Sato2017}
which at the same time possess a gapped bulk state and a gapless surface state, 
opened additional avenues towards a realization of these collective quantum phenomena
and their potential application in quantum computation.\cite{Nayak2008}  
In this context zero energy Majorana bound states (MBS), 
which represent the most simple non-Abelian excitation of Moore-Read states,\cite{Read2000} 
are particularly auspicious as they would allow the non-local storage of quantum bits, 
thereby promising a larger robustness against local sources of decoherence.\cite{Stone2006,Akhmerov2009}  

Various routes towards the realization of topological superconductors have been pursued. 
For example, it has been theoretically proposed that the proximity of an ordinary $s$-wave superconductor 
with a strong TI results in $p_x + ip_y$ superconductivity which can support MBS in vortices.\cite{Fu2008} 
Indeed, planar heterostructure could successfully be prepared by the epitaxial growth 
of topological insulators epitaxially grown on superconducting NbSe$_2$.\cite{Wang2012, Dai2017} 
In agreement with theoretical expectations an in-gap zero-bias conductance peak 
was found in magnetic vortices in proximity-coupled Bi$_2$Te$_3$ films.\cite{Xu2015}  

Another route may be the self-organized growth or atom-by-atom assembly 
of one-dimensional magnetic chains on strongly spin-orbit--coupled superconductors.  
Model calculations indicated that single-atomic chains with a modulated (helical) spin structure 
exhibit a non-trivial topological ground state with MBS at the two chain termination points.\cite{NadjPerge2013} 
Although a zero-bias conductance peak was indeed observed in scanning tunneling spectroscopy (STS) 
experiments performed on self-organized Fe chains on Pb(110),\cite{NadjPerge2014}
the interpretation of these results remains controversial.\cite{Ruby2015}  

\begin{table}[b]
\caption{Overview of experimental studies analyzing potential superconducting-doped topological insulators. 
The critical field $\mu_{0} H_{\rm C}$ was obtained for applied fields perpendicular to the sample surface.}
\vspace{0.5cm}
\begin{tabular}{p{3.0cm}p{5.3cm}p{2.0cm}p{3.0cm}p{1.7cm}p{0.6cm}}\hline\hline
Material (concentration) & Experimental techniques & $T_{\rm C}$ & $\mu_{0} H_{\rm C}$ & TSS verified? & Ref. \\ \hline
Cu$_{x}$Bi$_{2}$Se$_{3}$ & & & & \\
($0.10 \le x \le 0.15$) & transport, XRD, TEM, STM & 3.8\,K & 1.7\,T & No & [\onlinecite{Hor2010}]\\
($x = 0.25$) & transport, magnetometry & 3.3\,K & not shown & Yes & [\onlinecite{Lawson2012}]\\
($x = 0.3$) & point-contact spectroscopy & 3.2\,K & not shown & No & [\onlinecite{Sasaki2011}]\\
($x = 0.2$) & STM, STS & not shown & 1.7\,T (at 0.95\,K) & No & [\onlinecite{Levy2013}]\\
($x = 0.3$) & NMR & 3.4\,K & not shown & No & [\onlinecite{Matano2016}]\\
($x = 0.3$) & specific heat & 3.2\,K & not shown & No & [\onlinecite{Yonezawa2017}]\\
\hline
Sr$_{x}$Bi$_{2}$Se$_{3}$ & & & & \\
($x \leq 0.065$) & transport & 2.57\,K & $\approx$ 1\,T & Yes & [\onlinecite{Liu2015}]\\
($x = 0.1$) & transport, SEM, TEM & 2.9\,K & 1.4\,T (at 0\,K) & No & [\onlinecite{Shruti2015}]\\
($x = 0.2$) & STM, STS & 5\,K & $\ge$ 5\,T & Yes & [\onlinecite{Du2017}]\\
\hline
Tl$_{x}$Bi$_{2}$Te$_{3}$ & & & & \\
($x = 0.6$) & transport & 2.28\,K & 1.06\,T & No & [\onlinecite{Wang2016}]\\
($x = 0.5$) & ARPES & not shown & not shown & Yes & [\onlinecite{Trang2016}]\\
\hline
Nb$_{x}$Bi$_{2}$Se$_{3}$ & & & & \\
($x = 0.25$) & transport, STM, ARPES & 3.6\,K & 0.15\,T (at 2\,K) & Yes & [\onlinecite{Qiu2015}]\\
($x = 0.25$) & torque magnetometry & 3\,K & 0.6\,T & No & [\onlinecite{Asaba2017}]\\
\hline\hline
\end{tabular}
\label{tab_materials}
\end{table}
Especially from a materials perspective the intercalation or doping 
of bismuth chalcogenides (Bi$_2$Se$_3$ or Bi$_2$Te$_3$) represents another promising 
and frequently pursued approach towards the realization of topological superconductors. 
Throughout the remainder of this contribution, the term ``intercalation'' will refer 
to the inclusion of atoms into a van der Waals gap between two layers of the host material, 
whereas the term ``doping'' refers to impurities in regular lattice sites.  
Table\,\ref{tab_materials} summarizes experimental key results 
of materials combinations relevant in the context of our work. 
The left column of Tab.\,\ref{tab_materials} lists the chemical formula 
together with the nominal concentration of the doping element.
The following three columns, from left to right, recapitulate the employed experimental techniques, 
the reported critical temperature $T_{\rm C}$ and the critical field $\mu_{0} H_{\rm C}$, respectively. 
The second column from the right indicates whether the existence 
of the topological surface state (TSS) has explicitly been proven experimentally.  
We would like to emphasize that all measurements were either carried out at the Fermi level 
or it was shown that the TSS crosses $E_\textrm{F}$, a crucial condition towards topological superconductivity.
Finally, the column at the very right cites the reference.   

Cu$_x$Bi$_2$Se$_3$ with $0.10 \le x \le 0.15$ was the first intercalated topological material 
for which a superconducting transition with a transition temperature of 3.8\,K was claimed.\cite{Hor2010} 
X-ray diffraction (XRD) experiments in combination with transmission electron microscopy (TEM) 
confirmed that Cu was intercalated into the van der Waals gap 
of Bi$_2$Se$_3$ with ``long or short range order''.\cite{Hor2010} 
Torque magnetometry measurements that showed pronounced quantum oscillations 
in high magnetic field were interpreted as evidence for topological properties, 
although at a slightly different doping level ($x = 0.25$).\cite{Lawson2012} 
Furthermore, the observation of a zero-bias conduction peak in point-contact spectroscopy experiments 
which vanishes above 1.15\,K and 0.8\,T indicated the existence of Majorana fermions.\cite{Sasaki2011} 

An STM/STS study performed on Cu$_x$Bi$_2$Se$_3$ with $x = 0.2$ revealed 
an inhomogeneous sample that only exhibits superconductivity in some surface regions.\cite{Levy2013} 
Also other studies reported a rather low superconducting volume fraction 
which could be improved by avoiding substitutional Cu defects.\cite{Kriener2011}
In the superconducting regions of cleaved Cu$_{0.2}$Bi$_2$Se$_3$ 
the STM/STS study of Levy\,{\em et\,al.}\cite{Levy2013} mostly showed a gap without any zero-bias anomaly 
which could well be fitted by BCS theory, suggesting classical $s$-wave superconductivity.
Interestingly, some results occasionally obtained at particularly low tunneling resistance 
(close tip--sample distance) exhibited a zero-bias peak.  
Similar to what we will discuss below, this observation was ascribed to a tip which 
``became contaminated and possibly superconducting after crashing into the sample''.\cite{Levy2013} 
However, the study of Levy\,{\em et\,al.}\cite{Levy2013} does not discuss 
whether the superconducting regions also support the topological surface state.
Contradictory to the pairing mechanism suggested by these STS data, 
nuclear magnetic resonance (NMR) \cite{Matano2016} and specific heat \cite{Yonezawa2017} measurements demonstrated 
a broken spin-rotation symmetry indicating a pseudo spin-triplet state in Cu$_{x}$Bi$_{2}$Se$_{3}$.

Sr-intercalated and Sr-doped Bi$_2$Se$_3$ was investigated by resistivity experiments.  
Whereas no superconductivity was found for the doped sample, 
a superconducting transition temperature of 2.57\,K was reported for Sr-intercalated Bi$_2$Se$_3$. 
Magnetic field-dependent measurements revealed a critical field of roughly 1\,T 
and quantum oscillations verifying the presence of a surface state.\cite{Liu2015} 
In another transport study the influence of the nominal Sr content $x$ was analyzed 
and $x = 0.1$ was declared as the optimal level leading to a critical temperature of 2.9\,K 
and a critical field of 1.4\,T if interpolated to zero temperature.\cite{Shruti2015} 
In combined STM/STS measurements on Sr$_{0.2}$Bi$_2$Se$_3$ 
two kinds of surface areas were found.\cite{Du2017}
Whereas the clean surface with large atomically smooth terraces showed no signs of superconductivity, 
sample areas decorated with clusters of varying size revealed a V-shaped gap. 
Based on Landau level spectroscopy it was claimed that a proper topological surface state is present, 
even though the very narrow energy range over which the Landau levels could be observed 
severely complicates the determination if the Landau level dispersion relation 
indeed follows a Dirac-like behavior.\cite{Du2017}
Furthermore, the values reported in Ref.\,\onlinecite{Du2017} for the critical temperature ($T_{\rm C} = 5$\,K) 
and the critical field ($\mu_{0} H_{\rm C} \ge 5$\,T) by far exceed what was found 
in above-mentioned studies.\cite{Liu2015,Shruti2015} 

Recently, superconductivity with a critical temperature of 2.28\,K and critical field of 1.06\,T 
was found in Tl$_{x}$Bi$_2$Te$_3$ at optimal Tl content $x = 0.6$.\cite{Wang2016} 
X-ray diffraction showed sharp reflections similar to pristine Bi$_2$Te$_3$ indicating a good crystalline quality. 
An angle-resolved photoemission spectroscopy (ARPES) study of Tl$_x$Bi$_2$Te$_3$ 
with a slightly lower Tl content ($x = 0.5$) showed that the Dirac surface state is well isolated from bulk bands 
making this sample ideal to study topological superconductivity.\cite{Trang2016} 

Also Nb intercalation of Bi$_2$Se$_3$ leads to superconductivity 
with a critical temperature of 3.6\,K and a critical field of 0.15\,T at 2\,K. 
ARPES clearly demonstrated the existence of a Dirac-like surface state 
with a Dirac point energy lying roughly 300\,mV below the Fermi level.\cite{Qiu2015} 
Torque magnetometry showed a strong coupling between superconductivity and the crystal symmetry 
leading to bulk nematic order in the superconducting ground state\cite{Asaba2017} 
similar to the findings in Cu$_{x}$Bi$_{2}$Se$_{3}$. \cite{Yonezawa2017,Matano2016}

Although a large number of reports discusses the potential existence 
of topological superconductivity in doped or intercalated topological host materials,%
\cite{Hor2010,Lawson2012,Sasaki2011,Levy2013,Liu2015,Shruti2015,%
Du2017,Wang2016,Trang2016,Qiu2015,Asaba2017}
we have to conclude that none of these references at the same time convincingly proves 
the {\em coexistence} of surface superconductivity and a topological surface state with a Dirac-like dispersion.  
Most transport studies\cite{Hor2010,Sasaki2011,Liu2015,Shruti2015,Wang2016,Qiu2015,Asaba2017}
are essentially bulk sensitive, i.e., even a large volume fraction cannot safely assure 
that superconductivity extends all the way up to the terminating quintuple layer 
which supports the topological surface state.  
Others show indisputable evidence for surface superconductivity 
but fail to show the existence of the topological surface state (TSS).\cite{Levy2013}
Only one study claims the observation of surface superconductivity and the TSS.\cite{Liu2015}
However, even though the very high critical field of 5\,T suggests a type-II superconductor 
no Abrokosov lattice is shown,\cite{Liu2015} thereby questioning the validity of these results.
  
Here we report on our attempts to verify the coexistence of topological surface states 
and surface superconductivity on cleaved surfaces of Tl$_{0.6}$Bi$_2$Te$_3$ and Nb$_{0.25}$Bi$_2$Se$_3$. 
Especially Tl-doped Bi$_2$Te$_3$ is unique as---contrary to other materials---it is p-doped.  
In heavily n-doped materials the Fermi level is well inside the conduction band 
such that bulk and surface superconductivity may easily be confused.  
In contrast, we can disentangle topological states from bulk electronic properties for p-type Tl-doped Bi$_2$Te$_3$. 
Landau level spectroscopy in combination with quasiparticle interference mapping on one hand 
clearly demonstrates the presence of a topological surface state in both compounds. 
On the other hand, in neither case we were able to find clear evidence for surface superconductivity 
down to the lowest possible measurement temperature of 300\,mK.  
Even though superconducting gaps were regularly observed in STS data at high energy resolution, 
several inconsistencies indicate that these results are most likely caused 
by an unintentional coating of the tip apex with superconducting material, possibly in form of a small cluster. 
This interpretation is supported by values for the critical temperature and the critical field 
which deviated strongly from corresponding values measured with bulk sensitive methods.

\section{Experimental procedures}
\label{Sec:Experimental procedures}

The growth of Tl-doped Bi$_2$Te$_3$ was performed by mixing high-purity elemental shots 
of Tl (99.99\%), Bi (99.9999\%), and Te (99.9999\%) that were cleaned in order to remove oxide layers. 
The mixture was sealed in an evacuated quartz tube and heated to 1123\,K for 48\,h. 
The tube was cooled to 823\,K with a rate of 5\,K/h, followed by a quench in ice-water 
[see Refs.\,\onlinecite{Wang2015a,Wang2016} for further details].
Due to the high room temperature mobility of Tl atoms in Tl$_{x}$Bi$_2$Te$_3$ 
and the associated reduction of the superconducting volume fraction,\citep{Wang2016} 
samples were stored at liquid nitrogen temperatures or below. 

Scanning tunneling microscopy (STM) and spectroscopy (STS) experiments were carried out 
with two scanning tunneling microscopes, each covering a dedicated temperature range.  
Measurements with a minimal temperature $T_{\rm min} = 1.5\,\textrm{K}$ 
were performed with a home-built cryogenic scanning tunneling microscope 
which is equipped with a superconducting magnet that supplies a magnetic field 
of up to 12.5\,T perpendicular to the sample surface. 
Experiments at lower temperature ($T = 300$\,mK) 
were executed with a commercial Unisoku USM-1300. 
For all measurements we used electro-chemically etched W tips.  
Topographic images were recorded in the constant-current mode. 
The bias voltage is applied to the sample, i.e., negative (positive) voltages 
correspond to occupied (unoccupied) sample states, respectively.  
Spectroscopy data were obtained under open feedback loop conditions 
by adding a small modulation voltage $U_{\rm mod}$ 
to the bias voltage $U$ by means of a lock-in amplifier.

The samples were glued on a sample holder, 
introduced into the ultra-high vacuum system via a load lock, 
and cleaved at room temperature at a pressure $p \le 5 \cdot 10^{-10}$\,mbar.  
Immediately after cleavage the pristine sample was inserted into the STM. 
During tip conditioning, which was performed by dipping the tip apex into a clean Ag(111) surface, 
the sample was stored in a garage inside the LHe shield at temperatures close to 4.2\,K.

\section{Results and discussion}
\subsection{Tl-doped Bi$_2$Te$_3$} \label{Sec:Tl_Bi2Te3_topo}
\subsubsection{The topological surface state}

\begin{figure}[t]   
	\begin{minipage}[t]{0.55\textwidth} 
		\includegraphics[width=\columnwidth]{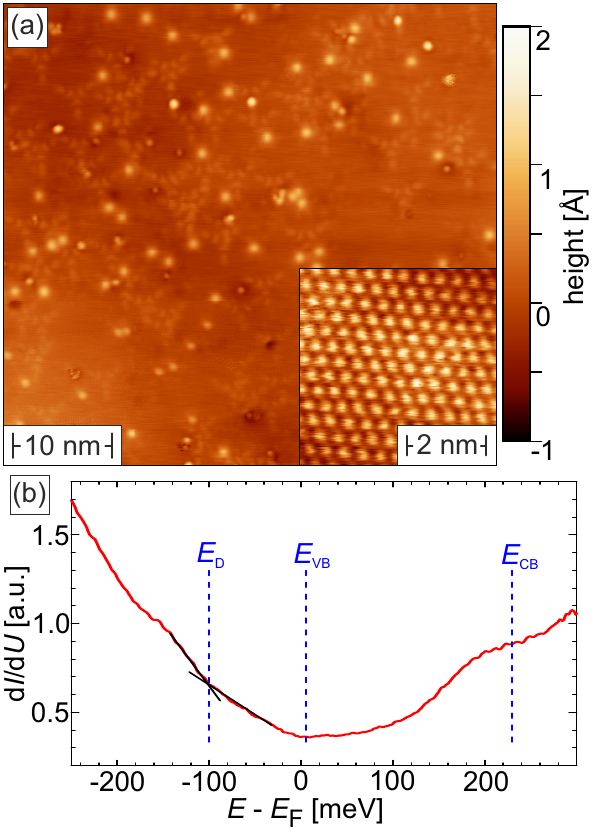}%
	\end{minipage}
	\hfill
	\begin{minipage}[b]{0.42\textwidth}
		\caption{(a) Topographic image of cleaved Tl$_{0.6}$Bi$_2$Te$_3$. 
		Defects characteristic for p-doped Bi$_2$Te$_3$ can be recognized. 
		No signs of Tl intercalation are visible. 
		The inset shows an atomic resolution image. 
		(b) Differential conductance (d$I/$d$U$) spectrum of Tl$_{0.6}$Bi$_2$Te$_3$
		recorded on the clean surface at a lateral distance of more than $3$\,nm from a defect. 
		Dirac point energy ($E_{\rm D}$), valence band maximum ($E_{\rm VB}$), 
		and conduction band minimum ($E_{\rm CB}$) are marked with blue dashed lines. 
		Scan/stabilization parameters: $T = 1.7$\,K; 
		(a) $U = -0.3\,$V, $I = 50\,$pA, (inset) $U = -0.2\,$V, $I = 50\,$pA;  
		(b) $U_{\rm set} = -0.4\,$V, $I_{\rm set} = 100\,$pA, $U_{\text{mod}} = 5\,$mV.} 
		 \label{Fig:Tl_Bi2Te3_topo}
	\end{minipage}	
\end{figure}

Figure\,\ref{Fig:Tl_Bi2Te3_topo}(a) shows the topography 
of a cleaved Tl$_{0.6}$Bi$_2$Te$_3$ sample.  
The inset presents a zoomed-in STM image with atomic resolution. 
The measured lattice constant of (4.4 $\pm$ 0.2)\,{\AA} 
is in good agreement with the bulk lattice constant of 4.38\,\AA. 
In a $50 \times 50$\,nm$^2$ scan area we find a total of $(220 \pm 15)$ defects,
i.e., far below the number of 9,000 Tl atoms expected 
within the first quintuple layer of Tl$_{0.6}$Bi$_2$Te$_3$.   
In fact, by analyzing the appearance of defects in STM images we find that the vast majority of them 
closely resemble defects which are also characteristic for pristine p-doped Bi$_2$Te$_3$, 
i.e., antisites of Bi in the top and the bottom Te plane of the first quintuple layer.\citep{Bathon2016}   
Based on these data one would naively expect that Tl does not occupy sites 
within Bi$_2$Te$_3$ quintuple layers but is rather located in van der Waals gaps between quintuple layers. 
This interpretation is, however, inconsistent with earlier neutron scattering and X-ray diffraction studies  
where an intercalated site could be excluded.\cite{Wang2016} 
Instead, these data indicated that the Tl atoms primarily occupy Bi sites, 
whereas the kicked-out Bi atoms in turn result in Bi$_{\rm Te}$ antisites, 
such that the system as a whole remains in the tetradymite structure. 
Effectively, the chemical formula of Tl$_{0.6}$Bi$_2$Te$_3$ 
approximately becomes (Tl$_{0.27}$Bi$_{0.73}$)$_{2}$(Bi$_{0.11}$Tl$_{0.89}$)$_{3}$. 
We can only speculate why Bi antisites, which are well visible 
on cleaved Bi$_2$Te$_3$,\citep{Bathon2016} might have remained undetected 
in our STM experiments on Tl$_{0.6}$Bi$_2$Te$_3$. 
Potentially, the bias voltage and, therefore, the energy window chosen here 
was unsuitable for detecting these antisites in topography. 

Figure\,\ref{Fig:Tl_Bi2Te3_topo}(b) presents 
an overview scanning tunneling spectrum of Tl$_{0.6}$Bi$_2$Te$_3$. 
Similar to the results presented in Ref.\,\onlinecite{Alpichshev2010}, 
the Landau level spectroscopy data which will be described below can consistently be explained if we assume that
the top of the bulk valence band is energetically located at the minimum differential conductance d$I/$d$U$.  
Correspondingly, the flat part of the shoulder in the unoccupied energy range 
marks the bottom of the conduction band at $E_{\rm CB} = (230 \pm 10)$\,meV. 
Similar to pristine Bi$_2$Te$_3$, the Dirac point lies inside the valence band 
and can be recognized by a change of slope in the d$I/$d$U$ signal. 
As indicated by black lines in Fig.\,\ref{Fig:Tl_Bi2Te3_topo}(b) this linear approximation procedure 
results in a Dirac energy $E_{\rm D} = (-100 \pm 20)$\,mV.

\begin{figure}[t]   
	\includegraphics[width=0.7\columnwidth]{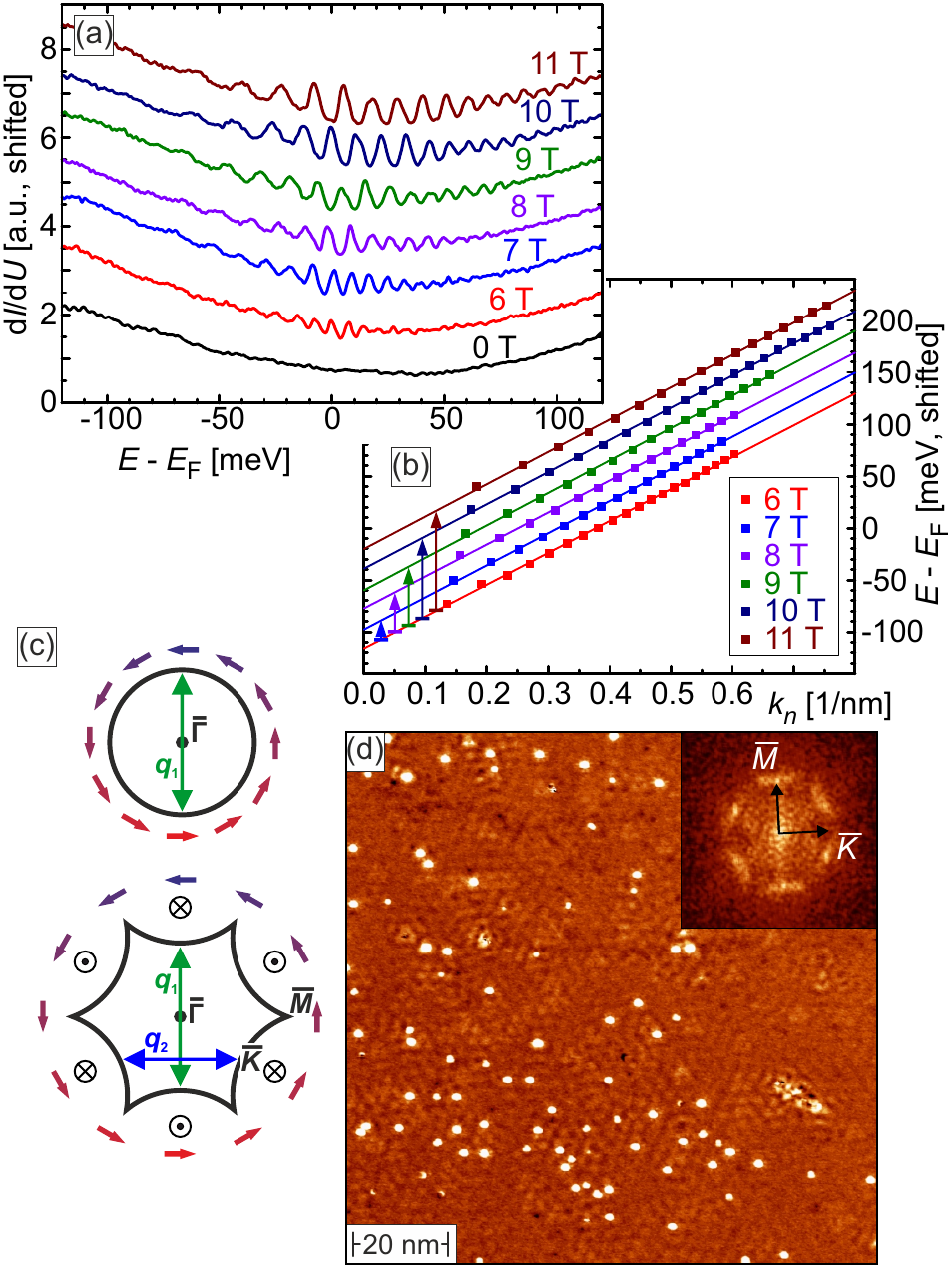}%
	\caption{(a) STS data of Tl$_{0.6}$Bi$_2$Te$_3$ measured at various magnetic fields at $T = 1.7$\,K. 
		(b) Landau level energies plotted versus $k_n$ confirming the linear dependence expected for a TSS. 
		(c) Schematic equipotential surface (black), in-plane spin polarization (blue/red arrows), 
		and potential scattering vectors ($\boldsymbol{q_{1}}$/$\boldsymbol{q_{2}}$) 
		for a TI close to the Dirac point (upper panel) and in the warped energy regime (lower). 
		(d) QPI map of Tl$_{0.6}$Bi$_2$Te$_3$ and its Fourier transformation in the inset ($T = 4.8$\,K). 
		No backscattering can be recognized. 
		Scan parameters: (a) $U = -150\,$mV, $I = 600\,$pA, $U_{\text{mod}} = 1\,$mV; 
		(d) $U = 400\,$mV, $I = 50\,$pA, $U_{\text{mod}}=10$\,mV.}   
	\label{Fig:Tl_Bi2Te3_LL_QPI}
\end{figure}   
Whereas the procedures performed to analyze the data of Fig.\,\ref{Fig:Tl_Bi2Te3_topo}(b) 
result in a relatively large error bar, Landau level spectroscopy (LLS) 
permits a much more accurate determination of the dispersion relation of Dirac-like quasiparticles.  
Fig.\,\ref{Fig:Tl_Bi2Te3_LL_QPI}(a) shows STS data of Tl$_{0.6}$Bi$_2$Te$_3$ 
measured at zero field (black) and at various externally applied magnetic fields up to 11\,T. 
The zero-field spectrum is similar to the one presented in Fig.\,\ref{Fig:Tl_Bi2Te3_topo}(b).  
At high magnetic fields a sequence of peaks appear in the energy range of $\pm$100\,meV. 
When increasing the magnetic field strength the individual peaks become more intense 
and their energy separation increases, such that more peaks become visible in a wider and wider energy range.  

These peaks indicate the evolution of a series of Landau levels (LL). 
In Dirac materials the LL energies are given by the equation 
\begin{equation}   
	E_{n} = E_{\rm D} +\textrm{sgn}(n) v_{\rm D} \cdot \sqrt{2\hbar e \vert n \vert B} 
		  = E_{\rm D} + \textrm{sgn}(n) v_{\rm D} \hbar k_{n} ,    \label{Eq:LLdispersion}
\end{equation}   
\noindent where $E_{\rm D}$ is the Dirac energy, $v_{\rm D}$ the Dirac velocity, 
$n$ the LL index, and $B$ the magnetic field perpendicular to the sample surface.
In the case of Tl$_{0.6}$Bi$_2$Te$_3$ all peaks visible in Fig.\,\ref{Fig:Tl_Bi2Te3_LL_QPI}(a) 
correspond to electron-like Landau levels; hole-like Landau levels could not be detected, 
in agreement with earlier experiments on pristine Bi$_2$Te$_3$.\citep{Okada2012} 
For further analysis we performed a background subtraction for the data 
presented in Fig.\,\ref{Fig:Tl_Bi2Te3_LL_QPI}(a) and fitted the Landau level peaks by Gaussian functions. 
The obtained Landau level energies are plotted versus $k_{n}$ in Fig.\,\ref{Fig:Tl_Bi2Te3_LL_QPI}(b). 
The results nicely match the expected linear dependence, 
thereby confirming the existence of a Dirac like surface state. 
Fitting the experimental data with Eq.\,(\ref{Eq:LLdispersion}) 
results in a Dirac point energy $E_{\textrm{D}} = -(118 \pm 1)$\,meV 
and a Dirac velocity $v_{\textrm{D}} = (4.7 \pm 0.1)\cdot 10^{5}$\,m/s.  

Another characteristic property of topological surface states is the absence of backscattering. 
It has been shown that quasiparticle interference mapping can visualize the scattering behavior 
of topological surface states.\cite{Roushan2009,PhysRevB.88.161407,Sessi2014} 
Fig.\,\ref{Fig:Tl_Bi2Te3_LL_QPI}(c) illustrates potential scattering vectors between nested parts 
of the equipotential surface of a TI in close proximity to the Dirac energy (upper panel) 
and further away from the Dirac point (lower panel),
where the symmetry of the crystal lattice leads to warping 
with a significant out-of-plane spin polarization caused by spin-orbit coupling.\cite{PhysRevLett.103.266801} 
Close to the Dirac energy the only nested (parallel) parts of the almost circular equipotential surface 
are located at opposite $k$ values, i.e., states which also possess an opposite spin polarization. 
Consequently, backscattering [see scattering vector $\boldsymbol{q_{1}}$ in Fig.\,\ref{Fig:Tl_Bi2Te3_LL_QPI}(c)] 
is forbidden as long as time-reversal symmetry is maintained.   
With increasing energy separation from the Dirac point spin-orbit effects which become more relevant,  
eventually deform the the equipotential surface to a snowflake-like shape. 
This opens further scattering channels ($\boldsymbol{q_{2}}$) which are allowed 
due to the resulting out-of-plane component to the spin polarization and expected to occur 
in the $\overline{\Gamma M}$ direction of the surface Brillouin zone.\cite{PhysRevLett.103.266801}   
Indeed, Fourier transformation of the quasiparticle interference map shown in Fig.\,\ref{Fig:Tl_Bi2Te3_LL_QPI}(d) 
which was measured at a bias voltage $U = 400\,$mV, i.e., at an energy about 0.5\,eV above the Dirac point,
shows intensity maxima along this direction, whereas no intensity is found 
along the backscattering direction $\overline{\Gamma K}$ (see inset). 

\subsubsection{Superconducting properties}
The results presented so far confirm that the surface of Tl-doped Bi$_2$Te$_3$ 
exhibits a topological surface state with a Dirac point about 100\,meV below the Fermi level 
and a valence band maximum just above the Fermi level.
These values slightly deviate from what has been observed by Trang et al.\,\cite{Trang2016} 
for as-grown (fresh) Tl$_{0.6}$Bi$_2$Te$_3$ by ARPES ($E_{\textrm{D}} = -60$\,meV). 
In the same study\cite{Trang2016} it was shown that sample storage 
at a pressure $p = 2 \times 10^{-10}$\,torr for 12\,h results in surface aging-induced n-type shift 
of the surface chemical potential by about 220\,meV, i.e., far into the conduction band. 
Even though the shift of 58\,mV observed here is much smaller, some surface band bending 
is undeniably observed in our STS data, the origin of which remains to be investigated.
We would like to emphasize, however, that the very low surface density of states at the Fermi level
is rather favorable for the unanimous identification of superconducting properties of Dirac electrons, 
as long as the surface band bending does not prevent to induce superconductivity from the bulk 
into the surface Dirac electrons via a proximity effect. 

\begin{figure}[t]
	\begin{minipage}[t]{0.48\textwidth} 
		\includegraphics[width=\columnwidth]{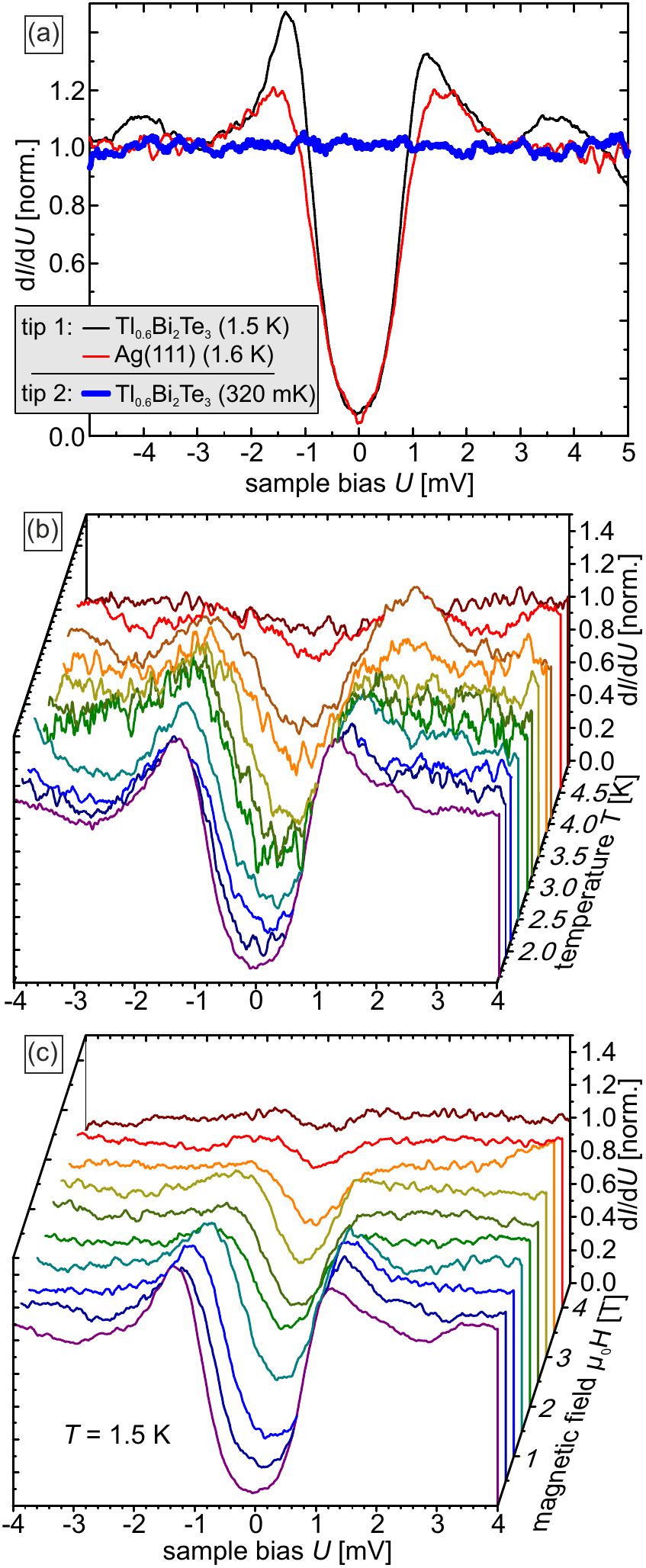}%
	\end{minipage}
	\hfill
	\begin{minipage}[b]{0.45\textwidth}
		\caption{(a) Normalized d$I$/d$U$ spectra of Tl$_{0.6}$Bi$_2$Te$_3$. 
		The spectrum taken with tip\,1 shows a superconducting gap at $T = 1.5$\,K (black line).  
		However, the same tip exhibits a very similar gap on Ag(111) (red), 
		indicating that a superconducting cluster was accidentally picked up, 
		presumably due to a gentle collision with the sample surface. 
		After careful tip handling even at $T = 0.32$\,K no gap can be found (tip 2; blue). 
		(b) Temperature-dependent d$I$/d$U$ spectra as measured with a superconducting tip.
		The gap vanishes at 5\,K, i.e., a temperature significantly higher than $T_{\rm C}$.  
		(c) Field-dependent STS data measured with the very same tip at $T = 1.5$\,K. 
		The superconducting gap can be recognized up to 4.5\,T. 
		Parameters: $U  =-5\,$mV, $I = 50 ... 200\,$pA, 
		$U_{\text{mod}} = 0.1\,$mV.}  \label{Fig:Tl_Bi2Te3_SC}
	\vspace{2cm}  \end{minipage}
\end{figure}
To investigate potential superconducting properties of cleaved Tl$_{0.6}$Bi$_2$Te$_3$  
we performed high resolution spectroscopy experiments close to the Fermi energy. 
An experimental result obtained at a nominal temperature $T \approx 1.5$\,K 
is presented as a black line in Fig.\,\ref{Fig:Tl_Bi2Te3_SC}(a).  
A U-shaped superconducting gap with an almost vanishing zero-bias d$I$/d$U$ signal 
and pronounced coherence peaks at $E \approx 1.3$\,meV is clearly visible. 
As shown in Fig.\,\ref{Fig:Tl_Bi2Te3_SC}(b), the gap size decreases 
with increasing temperature until it vanishes roughly at 5.0\,K, 
a value which is more than twice the reported critical temperature 
of bulk Tl$_{0.6}$Bi$_2$Te$_3$, $T_{\textrm{C}}^{\textrm{bulk}} = 2.3$\,K.\citep{Wang2016}
A similar discrepancy is found in the field-dependent STS data displayed in Fig.\,\ref{Fig:Tl_Bi2Te3_SC}(c).  
In qualitative agreement with the expected behavior the gap size decreases with increasing magnetic field. 
However, even at 4.5\,T a tiny gap can still be recognized. 
Similar to the temperature dependence the critical magnetic field, 
if it was extracted from the experimental data presented in Fig.\,\ref{Fig:Tl_Bi2Te3_SC}(c),
would substantially exceed the respective value of bulk Tl$_{0.6}$Bi$_2$Te$_3$, 
which only amounts to $\mu_{0} H_{\textrm{C}}^{\textrm{bulk}} \approx 400$\,mT 
at our measurement temperature of 1.5\,K.\cite{Wang2016}

We would like to emphasize that the observation of a superconducting gap in STS experiments 
on cleaved Tl$_{0.6}$Bi$_2$Te$_3$ was not at all an exception but was instead rather frequent, 
even if a freshly prepared W tip was used for the experiment.  
Furthermore, temperature- and field-dependent measurements reliably reproduced 
the data presented in Fig.\,\ref{Fig:Tl_Bi2Te3_SC}(b) and (c). 
Since the high magnetic field necessary to close the superconducting gap in Fig.\,\ref{Fig:Tl_Bi2Te3_SC}(c)
suggests a type-II superconductor, we expected an inhomogeneous phase 
where a sufficiently strong applied magnetic field ($H_{\rm app} > H_{\rm C_1}$) 
would lead to magnetic flux quanta penetrating the superconductor, potentially in form of an Abrikosov lattice.  
Although we have recently successfully observed Abrikosov lattices on Nb(110) (not shown) 
and the heavy-electron superconductor TlNi$_2$Se$_2$,\cite{Wilfert2018} 
we never found any magnetic vortices on cleaved Tl$_{0.6}$Bi$_2$Te$_3$.  

In order to characterize the probe tip more carefully, 
we performed STS experiments on a normal-metallic Ag(111) surface. 
The red curve in Fig.\,\ref{Fig:Tl_Bi2Te3_SC}(a) shows a high resolution spectrum 
recorded at $T = 1.6$\,K with a tip previously used for scanning a Tl$_{0.6}$Bi$_2$Te$_3$ sample. 
One can recognize a superconducting gap with a very similar shape 
as for Tl$_{0.6}$Bi$_2$Te$_3$ (cf.\ black spectrum). 
The small deviations are possibly caused by a slightly higher measurement temperature 
and the metallic nature of silver which leads to a much higher density of states at the Fermi level. 
These findings clearly show that the superconducting gap 
observed in the black curve of Fig.\,\ref{Fig:Tl_Bi2Te3_SC} 
is not caused by the sample itself but by the tip which most likely picked up 
a superconducting cluster from the Tl$_{0.6}$Bi$_2$Te$_3$ sample. 

Comparison of our results with a recent temperature- and field-dependent STS study\cite{Du2017}
of the potential topological superconductor Sr$_{0.2}$Bi$_2$Se$_3$ reveals some surprising similarities.  
For example, for Sr$_{0.2}$Bi$_2$Se$_3$ a critical temperature $T_{\rm C} \approx 5$\,K
and a second critical field $\mu_{0} H_{\rm C_2} \approx 5$\,T were reported.\cite{Du2017} 
These values do not only significantly exceed the corresponding bulk values 
determined by transport,\cite{Liu2015,Shruti2015} they are also strikingly close 
to what we observe with superconducting tips in Fig.\,\ref{Fig:Tl_Bi2Te3_SC}.  
Even though the existence of a second critical field would necessarily imply 
the formation of magnetic vortices at $H_{\rm C_1} < H < H_{\rm C_2}$, 
their existence has neither been discussed nor experimentally shown in Ref.\,\onlinecite{Du2017}.  

Only if we took extreme care to safely approach the tip to the Tl$_{0.6}$Bi$_2$Te$_3$ surface 
by using a very low set point current $I_{\textrm{set}}\leq 30$\,pA and feedback circuit settings 
which exclude self-resonance we could avoid any tip--sample contact.  
With these clean W tips we performed measurements on several samples 
that were previously analyzed by transport measurements.  
Even though all samples showed bulk superconductivity with transition temperatures close to 2.3\,K, 
we were unable to detect any superconducting gap in the low-bias regime around the Fermi level 
in our STS experiments, as exemplarily shown by the thick blue line in Fig.\,\ref{Fig:Tl_Bi2Te3_SC}(a). 
In combination with the previously discussed observations, i.e., the strong evidence 
for an (accidental) superconducting probe tip and the absence of flux quanta, 
these measurements exclude that the surface of cleaved Tl$_{0.6}$Bi$_2$Te$_3$ 
is superconducting down to the lowest measurement temperature applied here, i.e., $T = 0.32$\,K. 

This finding might be explained by three scenarios: 
(i) The sample was superconducting immediately after materials synthesis 
but later degrades due to aging effects, 
potentially because of insufficient cooling and the resulting onset of Tl segregation,
which could lead to a reduction of the superconducting volume fraction.\citep{Wang2016}
(ii) Superconductivity only exists in the bulk and the proximity effect on Dirac electrons 
is too weak to observe any significant Cooper pair formation on the sample surface.  
(iii) The surface critical temperature of Tl$_{0.6}$Bi$_2$Te$_3$ 
is substantially lower than the corresponding bulk value, $T_{\rm C}^{\rm bulk} = 2.3$\,K.

\begin{figure}[t]
\centering
\includegraphics[width=0.94\textwidth]{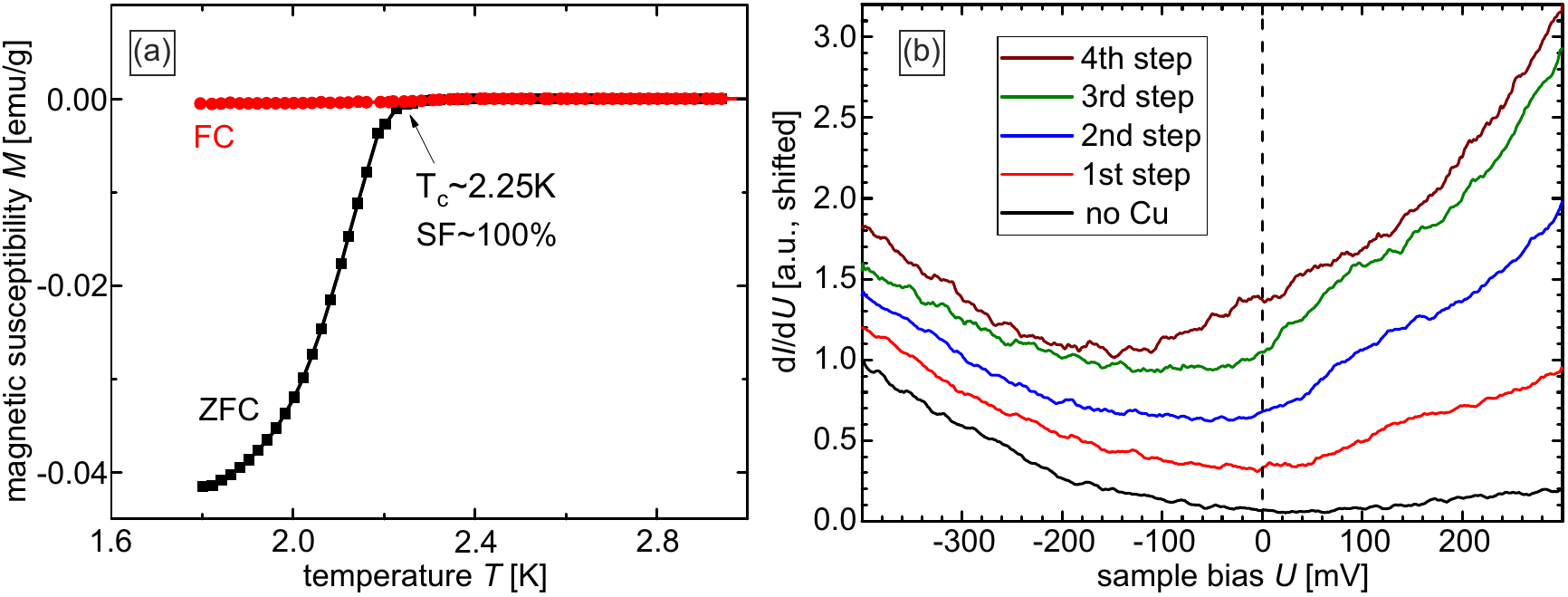}%
\caption{(a) Temperature-dependent magnetic susceptibility measured after STM/STS measurements. 
	No significant changes of critical temperature and the superconducting volume fraction were found.
	(b) Overview spectroscopy data taken with a W tip 
	on Tl$_{0.6}$Bi$_2$Te$_3$ for different Cu doping amounts. 
	With increasing amount of Cu on the sample surface the bulk band gap shifts to lower energies. }
\label{Fig:Susz_Cu_doping}
\end{figure}
To check if degradation, scenario (i), is responsible for the absence of surface superconductivity 
we performed another set of SQUID measurements after STS experiments had been carried out, 
i.e., the sample initially shipped from the growth lab (University of Cologne) 
to the STM lab (Universit\"{a}t W\"{u}rzburg) is, upon completion of STM measurements, shipped back to its origin.  
Fig.\,\ref{Fig:Susz_Cu_doping}(a) shows the resulting temperature-dependent magnetic susceptibility 
which clearly demonstrates that $T_{\rm C}^{\rm bulk}$ remains unchanged 
and that the superconducting volume fraction still amounts close to 100\%. 
Obviously, the small shift of the chemical potential is only present at the surface, 
whereas the bulk chemical potential remains unchanged throughout the entire experiment.

To test scenario (ii) we increased the density of bulk states around the Fermi energy 
to potentially boost the coupling between topological Dirac electrons on one hand 
and the Cooper pairs supported by the Tl$_{0.6}$Bi$_2$Te$_3$ bulk on the other hand. 
This goal was achieved by doping the Tl$_{0.6}$Bi$_2$Te$_3$ surface with Cu evaporated
onto the cold sample inside the LT-STM, which resulting in well separated Cu adatoms (not shown here). 
Non-magnetic Cu was chosen to avoid any influences of magnetic moments on the superconductivity. 
Fig.\,\ref{Fig:Susz_Cu_doping}(b) presents overview spectroscopy data 
measured on pristine Tl$_{0.6}$Bi$_2$Te$_3$ and after four consecutive steps of Cu adatoms deposition. 
With increasing Cu density we observed that the minimum of the d$I$/d$U$ signal shifts systematically to lower energies. 
Although we cannot directly detect the top and bottom edge of the valence and conduction band, respectively, 
it is quite reasonable to assume that the gap shifts accordingly.
Obviously, the adatoms donate electrons to Tl$_{0.6}$Bi$_2$Te$_3$ and lead to an n-type shift of the chemical potential, 
a behavior also observed for other $3d$ transition metals on topological insulators.\cite{Sessi2014}  
The bulk band gap shifts from the Fermi energy for the pristine sample by about 200\,meV towards the occupied states, 
eventually resulting in a Fermi level that lies well inside the conduction band after four steps of doping. 
In spite of this significant electron doping, we could not detect any sign of superconductivity 
in Cu-doped Tl$_{0.6}$Bi$_2$Te$_3$ down to $T = 1.5$\,K (high resolution spectra not shown here). 

The surface-related downward band bending effects discussed above 
possibly results in a surface which is effectively decoupled from the p-doped bulk. 
This decoupling could lead to scenario (iii), i.e., a surface critical temperature 
which is substantially lower than the corresponding bulk value.  
In fact we could not detect any sign of surface superconductivity in our STS measurements 
down the lowest possible sample temperature in our study 
which was more than a factor of 6 lower than $T_{\rm C}^{\rm bulk} = 2.3$\,K. 
As will be discussed below the issue of a p--n junction which isolates the surface from the bulk 
might be overcome by p-type doping the Tl$_{0.6}$Bi$_2$Te$_3$ surface.

\subsection{Nb-intecalated Bi$_2$Se$_3$} \label{Sec:Nb_Bi2Se3_topo}

In an attempt to potentially detect superconductivity 
on an intrinsically n-doped superconducting topological insulator, 
we also investigated the intercalation material Nb$_{x}$Bi$_2$Se$_3$ which exhibits 
a bulk critical temperature $T_{\rm c}^{\rm bulk} = 3.2$\,K.\cite{Qiu2015} 
Similar to what has been shown in Fig.\,\ref{Fig:Tl_Bi2Te3_LL_QPI}(b) for Tl$_{0.6}$Bi$_2$Te$_3$
we could also confirm the existence of a topological surface state for Nb$_{x}$Bi$_2$Se$_3$ 
based on the linear dispersion of the LL peaks in a strong external magnetic field, 
resulting in a Dirac point energy $E_{\textrm{D}} = -(350 \pm 2)$\,meV (not shown here). 
\begin{figure}[t]
\begin{minipage}[b]{0.48\textwidth} 
		\includegraphics[width=\columnwidth]{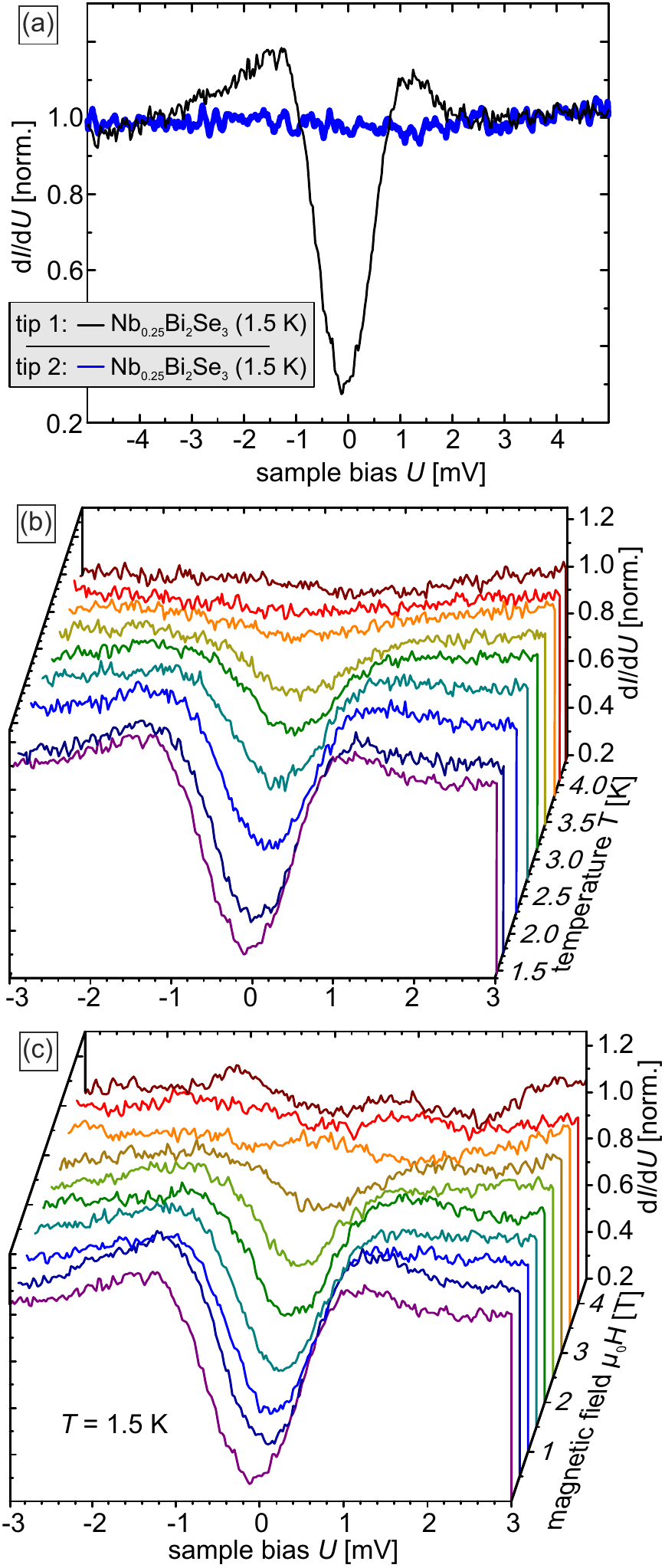}%
	\end{minipage}
	\hfill
	\begin{minipage}[b]{0.45\textwidth}
		\caption{(a) High resolution STS data on Nb-intercalated Bi$_2$Se$_3$ acquired with 
		a superconducting (1; black) and a normal-metallic W tip (2; blue) at a nominal temperature $T = 1.5$\,K. 
		While for tip\,1 a pronounced gap is visible no superconductivity can be recognized with tip\,2. 
		(b) Temperature-dependent normalized d$I$/d$U$ signal 
		taken on Nb$_{x}$Bi$_2$Se$_3$ with a superconducting tip. 
		(c) Magnetic field-dependent STS data taken with superconducting tip. 
		The gap can be recognized up to about 3\,T. 
		For higher fields additional variations caused by the fine structure of the Landau levels are visible. 
		Stabilization parameters: $U = -5\,$mV, $I = 200\,$pA, $U_{\text{mod}} = 0.1\,$mV.}  \label{Fig:NbBi2Se3_SC}
	\vspace{3cm}  \end{minipage}
\end{figure}
Regarding superconductivity, Fig.\,\ref{Fig:NbBi2Se3_SC}(a) shows high resolution STS data obtained with various tips.  
As already described in the context of Fig.\,\ref{Fig:Tl_Bi2Te3_SC} for Tl$_{0.6}$Bi$_2$Te$_3$, 
we also detected superconducting gaps on Nb$_{x}$Bi$_2$Se$_3$ with numerous tips, 
as shown by the spectrum colored black in Fig.\,\ref{Fig:NbBi2Se3_SC}(a). 
Temperature- and magnetic field-dependent d$I$/d$U$ spectra measured with such a tip 
are presented in Fig.\,\ref{Fig:NbBi2Se3_SC}(b) and (c), respectively. 
These data reveal that both the critical temperature and the critical field are very similar to the values 
detected with superconducting tips on Tl$_{0.6}$Bi$_2$Te$_3$ [cf.\ Fig.\,\ref{Fig:Tl_Bi2Te3_SC}(b) and (c)], 
whereby the critical magnetic field is substantially higher than the bulk value 
$\mu_{0} H_{\textrm{C}}^{\textrm{bulk}} \approx 200$\,mT reported for $T = 1.5$\,K.\cite{Qiu2015} 
Furthermore, we could not detect any hint of magnetic flux quanta or an Abrikosov lattice.
These observations together with the fact that we found no gap with some particularly carefully handled STM tips 
indicate that also for Nb$_{x}$Bi$_2$Se$_3$ the gap does not represent a property of the sample 
but is rather caused by a superconducting cluster which was unintentionally picked up by the tip.  

\subsection{Discussion}

Our investigation of the electronic properties of the doped topological insulators 
Tl$_{0.6}$Bi$_2$Te$_3$ and Nb$_{x}$Bi$_2$Se$_3$ reveals some striking similarities.  
Quasiparticle interference and Landau level spectroscopy (LLS) data confirm that the surfaces 
of both materials exhibit a linearly dispersing topological surface state which is protected from backscattering. 
Although magnetic susceptibility measurements undoubtedly confirm the Meissner effect, 
suggesting a superconducting volume fraction close to 100\%, 
our highly surface sensitive STS data indicate that the surface of both materials is normal conducting.   
Precise analysis of LLS data obtained on Tl$_{0.6}$Bi$_2$Te$_3$ reveals 
that---compared to earlier ARPES data\cite{Trang2016}---the surface is moderately n-shifted, 
placing the chemical potential right into the band gap where the density of states is minimal.  
Doping the surface with Cu adatoms resulted in a very significant downwards (n-type) surface band bending 
which---due to the rather low carrier density near the surface---will reach rather deep into the bulk.  
Potentially, this downward band bending leads to a further decoupling 
of the n-doped surface from the p-doped Tl$_{0.6}$Bi$_2$Te$_3$ bulk. 
One possibility to overcome the issue of a p--n junction which isolates the surface from the bulk 
might be through ``hole'' doping the Tl$_{0.6}$Bi$_2$Te$_3$ surface with a suitable atom/molecule, such as C$_{60}$.

Another result of our study is that even gentle collisions of the STM tip with any of the two materials investigated here, 
i.e., Tl$_{0.6}$Bi$_2$Te$_3$ or Nb$_{x}$Bi$_2$Se$_3$, lead to superconducting tips.  
For both materials, temperature- and field-dependent measurements reveal very similar values 
for the critical temperature and the critical field of $T_{\rm C} \approx 5$\,K 
and $\mu_{0} H_{\rm C_2} \approx 5$\,T, respectively.  
It appears that---even though both the surfaces of extended single crystals and the W tip 
are normal conducting themselves---the combinations of W plus the nanoparticles 
picked up by STM tip very reliably exhibit a superconducting gap which survives 
up to critical temperatures or external magnetic fields well above the corresponding bulk values. 
The large critical field could be due to a finite-size effect, but the enhancement in $T_{\rm C}$ 
cannot be explained in this way, thereby pointing to a new superconducting phase. 
In this regard, it is useful to mention that W has a tendency towards superconducting instability.\cite{BP1968}
Future dedicated studies on nanoparticles will be necessary to test this hypothesis.  

\section{Conclusion}

In conclusion, we analyzed the surface structural and electronic properties of Tl$_{0.6}$Bi$_2$Te$_3$ 
and Nb$_{x}$Bi$_2$Se$_3$ by low-temperature scanning tunneling spectroscopy. 
We could clearly demonstrate the existence of a topological surface state 
by Landau level spectroscopy and quasiparticle interference. 
The not-so-occasional observation of gapped tunneling spectra is ascribed to tips 
which were unintentionally coated with superconducting material.
In these cases both the critical temperature and the critical magnetic field 
are about a factor of 3--10 higher than in bulk sensitive experiments.  
In agreement with this hypothesis we could not find any hint 
of magnetic vortices in field-dependent measurements.  
We conclude that the topological surface state of cleaved Tl$_{0.6}$Bi$_2$Te$_3$ 
and Nb$_{x}$Bi$_2$Se$_3$ surfaces do not take part in the superconducting properties 
of these materials down to the lowest possible measurement temperatures of 300\,mK. 

\section*{Acknowledgments} 

The experimental work at Universit\"{a}t W\"{u}rzburg 
was supported by DFG (through SFB 1170 ``ToCoTronics''; project C02).
The work at Cologne has received funding from the European Research Council (ERC) 
under the European Union's Horizon 2020 research and innovation programme (grant agreement No 741121) 
and was also supported by DFG (CRC1238 ``Control and Dynamics of Quantum Materials'', Projects A04). 
Y.S.H. acknowledges support from the NSF DMR-1255607.
W.W. acknowledges support from the US NSF grant DMR-1506618.
M.C.M. and A.F.O. acknowledge support from the Netherlands Organisation for Scientific Research (NWO).



%

\end{document}